\documentclass[sigconf,natbib]{acmart}

\usepackage{enumitem}
\usepackage{hyperref}
\usepackage{multirow}
\usepackage{amsthm}
\usepackage{array}
\usepackage{subfigure}
\usepackage{graphicx}
\setlist[enumerate]{align=left}
\usepackage{algorithm}
\usepackage{algorithmic}
\usepackage{subcaption}

\usepackage[multiple]{footmisc}

\AtBeginDocument{%
  }

\copyrightyear{2026}
\acmYear{2026}
\setcopyright{cc}
\setcctype{by}
\acmConference[WSDM '26]{Proceedings of the Nineteenth ACM International Conference on Web Search and Data Mining}{February 22--26, 2026}{Boise, ID, USA}
\acmBooktitle{Proceedings of the Nineteenth ACM International Conference on Web Search and Data Mining (WSDM '26), February 22--26, 2026, Boise, ID, USA}
\acmPrice{}
\acmDOI{10.1145/3773966.3777992}
\acmISBN{979-8-4007-2292-9/2026/02}

\begin{document}

\title{Automated Information Flow Selection for Multi-scenario Multi-task Recommendation}

\author{Chaohua Yang}
\affiliation{%
  \institution{College of Computer Science and Software Engineering, Shenzhen University}
  \city{Shenzhen}
  \country{China}
}
\email{chaohua.ych@gmail.com}

\author{Dugang Liu}
\authornote{Co-Corresponding authors}
\affiliation{%
  \institution{College of Computer Science and Software Engineering, Shenzhen University}
  \city{Shenzhen}
  \country{China}}  
\email{dugang.ldg@gmail.com}

\author{Shiwei Li}
\affiliation{%
  \institution{Huazhong University of Science and Technology}
  \city{Wuhan}
  \country{China}
}
\email{lishiwei@hust.edu.cn}

\author{Yuwen Fu}
\affiliation{
  \institution{University of Chinese Academy of Sciences}
  \city{Shenzhen}
  \country{China}
}
\email{fuyuwen16@mails.ucas.ac.cn}

\author{Xing Tang}
\affiliation{
  \institution{Shenzhen Technology University}
  \city{Shenzhen}
  \country{China}
}
\email{xing.tang@hotmail.com}

\author{Weihong Luo}
\affiliation{
  \institution{Financial Technology (FIT), Tencent}
  \city{Shenzhen}
  \country{China}
}
\email{lobbyluo@tencent.com}

\author{Xiangyu Zhao}
\affiliation{%
  \institution{City University of Hong Kong}
  \city{Hongkong}
  \country{China}}
\email{xianzhao@cityu.edu.hk}

\author{Xiuqiang He}
\affiliation{%
  \institution{Shenzhen Technology University}
  \city{Shenzhen}
  \country{China}
}
\email{hexiuqiang@sztu.edu.cn}

\author{Zhong Ming}
\authornotemark[1]
\affiliation{%
  \institution{Shenzhen University}
  \city{Shenzhen}
  \country{China}
}
\email{mingz@szu.edu.cn}

\renewcommand{\shortauthors}{Chaohua Yang, et al.}

\begin{abstract}
Multi-scenario multi-task recommendation (MSMTR) systems must address recommendation demands across diverse scenarios while simultaneously optimizing multiple objectives, such as click-through rate and conversion rate. 
Existing MSMTR models typically consist of four information units: scenario-shared, scenario-specific, task-shared, and task-specific networks.
These units interact to generate four types of relationship information flows, directed from scenario-shared or scenario-specific networks to task-shared or task-specific networks. 
However, these models face two main limitations:
1) They often rely on complex architectures, such as mixture-of-experts (MoE) networks, which increase the complexity of information fusion, model size, and training cost.
2) They extract all available information flows without filtering out irrelevant or even harmful content, introducing potential noise.
Regarding these challenges, we propose a lightweight \textbf{Automated Information Flow Selection (AutoIFS)} framework for MSMTR. 
To tackle the first issue, AutoIFS incorporates low-rank adaptation (LoRA) to decouple the four information units, enabling more flexible and efficient information fusion with minimal parameter overhead.
To address the second issue, AutoIFS introduces an information flow selection network that automatically filters out invalid scenario-task information flows based on model performance feedback. 
It employs a simple yet effective pruning function to eliminate useless information flows, thereby enhancing the impact of key relationships and improving model performance. 
Finally, we evaluate AutoIFS and confirm its effectiveness through extensive experiments on two public benchmark datasets and an online A/B test.
\end{abstract}



\maketitle

\section{Introduction}\label{sec:intro}
Recommendation systems have become a key technical means to alleviate information overload by accurately capturing user preferences and quickly matching relevant information~\cite{2005toward, cheng2016wide, zhang2019deep, ko2022survey, tang2025retrieval, ouyang2021mobile}.
With the widespread application of recommendation systems, diverse business scenarios and user needs have become core challenges.
Traditional recommendation systems typically focus on a single objective within a specific scenario, such as click prediction in e-commerce~\cite{DeepFM, xdeepfm, DIN}.
However, industrial applications, such as online financial platforms, often require models to serve multiple scenarios (e.g., homepage, balanced investment page, aggressive investment page) and multiple tasks (e.g., click prediction, purchase prediction, purchase amount prediction).
These scenarios and tasks are typically interconnected in complex ways and may also involve potential conflicts.
Therefore, designing a unified multi-scenario multi-task recommendation (MSMTR) model to capture their intricate interactions efficiently has become a new research focus~\cite{m2m,3mn, hinet,m3oe, zou2022automatic, song2024multi}.

\begin{figure}[htbp]
    \centering
    \includegraphics[width=0.95\linewidth]{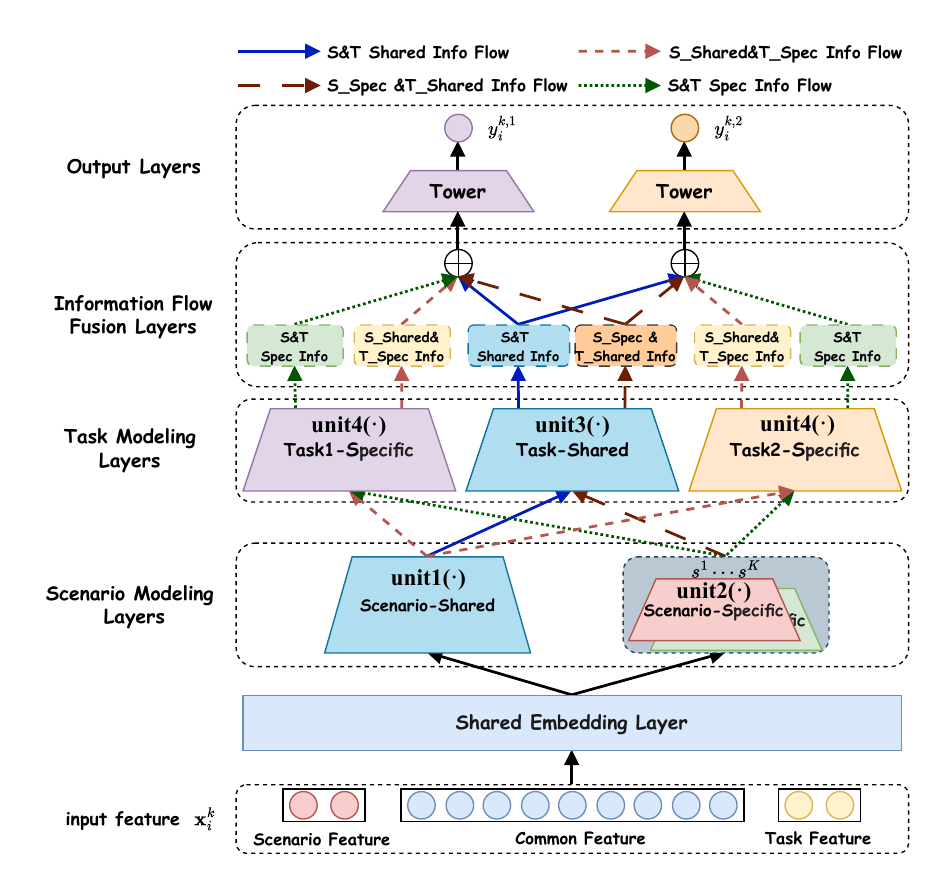}
    \Description[<Figure 1. Fully described in the text.>]{<A full description of Figure 1 can be found in the second paragraph of Section 1.>}
    \caption{Overview of the multi-scenario multi-task recommendation framework, where arrows of different colors represent different relationship information flows.}
    \label{fig:intro}
    \vspace{-5pt}
\end{figure}
The core of multi-scenario multi-task modeling lies in effectively integrating multi-scenario learning and multi-task learning while accurately characterizing the complex dependencies between scenarios and tasks.
Existing MSMTR models typically adopt a hierarchical modeling paradigm, whose architecture can be simplified as shown in Fig.~\ref{fig:intro}, where two tasks are used for illustration.
This architecture typically comprises an embedding layer, scenario modeling layers, task modeling layers, output layers, and a fusion operation that integrates scenarios and tasks. For clarity, this fusion can be abstracted as the information flow fusion layers in Fig.~\ref{fig:intro}.
Multi-scenario learning needs to consider both the commonalities and specifics of the scenarios~\cite{HCC, STAR, LLM4MSR}, and a similar consideration is also essential in multi-task learning~\cite{mtl, mmoe, tang2015predicting}.
Thus, in the scenario- and task-modeling layers of Fig.~\ref{fig:intro}, there are four types of information units: scenario-shared, scenario-specific, task-shared, and task-specific networks.
The interaction among these four units generates four distinct scenario-task relationship information flows, directed from scenario-shared or scenario-specific networks to task-shared or task-specific networks. These flows can be viewed as the decomposed components of the complex dependencies between scenarios and tasks.
Such complex dependencies imply that the importance of different relationship information should vary across scenarios and tasks.

Although existing MSMTR models can capture the four types of relationship information either implicitly or explicitly, they often fail to distinguish their distinct effects, instead fusing all information flows indiscriminately to generate the final representation.
However, some relationship information may be redundant or even harmful, hindering the modeling of critical dependencies and degrading overall performance.
To better capture the complex scenario-task interactions, it is therefore essential to adaptively select relevant information flows while filtering out redundant ones based on the needs of different scenarios and tasks.
In addition, existing MSMTR models often rely on complex architectures, such as mixture-of-experts (MoE) or multilayer perceptrons (MLPs), to explicitly characterize the four types of relational information. These architectures complicate information fusion and lead to substantial parameter growth as the number of scenarios and tasks increases, ultimately limiting optimization and efficiency.
Therefore, designing a lightweight architecture that explicitly models these four types of relationship information while enabling efficient and straightforward fusion remains an open and challenging problem.

To address the above challenges, this paper proposes an \textbf{Automated Information Flow Selection (AutoIFS)} framework for multi-scenario multi-task recommendation. 
AutoIFS features two main components: efficient modeling of four types of relationship information and adaptive selection of information flows from scenarios to tasks.
Specifically, to improve model efficiency and simplify information fusion, AutoIFS employs low-rank adaptation (LoRA) \cite{LoRA, li2025toplora, li2025bora} to decouple scenario-shared, scenario-specific, task-shared, and task-specific information.
It maintains a LoRA adapter for each task or scenario, achieving practical information decoupling while enabling more flexible and straightforward information fusion.
Furthermore, AutoIFS incorporates an information flow selection network that automatically identifies and selects crucial relational information between scenarios and tasks.
A simple pruning function is used to remove useless relationship information, preventing negative information flows and enhancing the contribution of key relationship information in modeling users’ valid preferences.
Finally, we extensively evaluated AutoIFS on two public benchmark datasets and validated its practical effectiveness through A/B testing in real-world business scenarios.
\section{Related Work}\label{sec:related}
This section briefly outlines representative work on two research topics: multi-scenario multi-task recommendation and low-rank adaptation in recommendation.

\noindent\textbf{Multi-Scenario Multi-Task Recommendation.}
Combining multi-scenario and multi-task learning to model the complex relationships between scenarios and tasks in real applications has attracted increasing attention in recent years~\cite{m2m, 3mn, pepnet, hinet, m3oe, song2024multi, zou2022automatic, ouyang2023masked, ding2024towards}.
Existing MSMTR models typically adopt a hierarchical modeling paradigm, which can be grouped into two main categories based on how they learn and integrate scenario and task information.
The first category is based on the MoE network architecture design to model scenario and task information~\cite{hinet, m3oe, zou2022automatic}.
For example, HiNet~\cite{hinet} employs a hierarchical extraction network to transfer valuable information across scenarios while preserving scenario- and task-specific features, whereas M3oE~\cite{m3oe} uses three MoE modules to learn common, domain-specific, and task-specific user preferences.
The second category is based on the dynamic weight generation mechanism~\cite{m2m, pepnet, 3mn}, utilizing meta-networks to incorporate scenario and task information separately.
For instance, M2M~\cite{m2m} designs a meta-attention and a meta-tower module to combine scenario knowledge and capture inter-scenario correlations, and 3MN~\cite{3mn} introduces three meta-networks to learn scenario-related knowledge and the mutual correlation between scenarios and tasks.
However, these methods focus on modeling the entire relationship space between scenarios and tasks through complex architectures.
In contrast, our method adopts a lightweight design and introduces an information flow selection network to eliminate unhelpful relationship information between scenarios and tasks adaptively.

\noindent\textbf{Low-Rank Adaptation in Recommendation.}
Low-rank adaptation (LoRA) is an efficient fine-tuning technique for large pre-trained models (LLMs)~\cite{LoRA, li2025beyond}, introducing low-rank matrices to learn task-specific information effectively.
In recent years, it has received increasing attention in deep recommender systems and has been applied to both model architecture design and fine-grained tuning of LLMs~\cite{mlora,multilora, reclora, ilora, zhao2024collaborative, qin2024atflrec}.
For model architecture design, MLoRA~\cite{mlora} introduces a dedicated LoRA module for each domain to capture domain-specific information, while MultiLoRA~\cite{multilora} proposes a multi-directional low-ranking adaptation paradigm for multi-domain recommendation, dividing domains in a fine-grained manner and modeling domain preferences.
For fine-grained tuning of LLMs, RecLoRA~\cite{reclora} introduces a personalized LoRA module that maintains a set of LoRA parameters for different users, enabling LLMs to learn users' lifelong personalized preferences, and iLoRA~\cite{ilora} designs instance-level LoRA to fine-tune LLMs for learning user sequential behavior representations.
Unlike these methods, our AutoIFS incorporates LoRA into multi-scenario multi-task recommendations, effectively decoupling the complex relationships between scenarios and tasks for more flexible modeling.

\section{Problem Formulation}\label{sec:problem}
This section defines the key concepts and notations for the MSMTR problem. 
For all scenarios and tasks, the feature space 
$\mathcal{X}$ consists of user, item, and context features, while the label space $\mathcal{Y}=\mathcal{Y}^1 \times \cdots \times \mathcal{Y}^M$ represents the interaction outcomes of $M$ tasks, where each task $\mathcal{Y}^m \in \mathcal{Y}$ is typically binary (e.g., \textit{interaction} or \textit{non-interaction}).
Given $K$ scenarios and $M$ tasks, the training instances in MSMTR can be denoted as $\mathcal{D}= \{(\mathbf{x}_i^k,\{y_i^{k,m}\}_{m=1}^M) \}_{k=1}^K$,
where $\mathbf{x}_i^k \in \mathcal{X}$ is the feature vector of the $i$-th instance in $k$-th scenario, and $y_i^{k,m} \in \{0,1\}$ is the label corresponding to the $k$-th scenario and the $m$-th task.
Based on the above training instances set, multi-scenario multi-task modeling aims to train a model $\hat{y}_i^{k, m}=\mathcal{F}(\mathbf{x}_i^k)$ to serve multiple scenarios and tasks
 ~\cite{m3oe}, where $\mathcal{F}(\cdot)$ is the mapping function from features to labels implemented by different models.
In practice, the cross-entropy function is commonly employed to optimize the model,
\begin{equation}\label{eq:1}
    \mathcal{L}=\sum_{m=1}^M\sum_{k=1}^K\sum_{i=1}^{|s^k|}l(y_i^{k,m}, \hat{y}_i^{k,m}),
\end{equation}
where $|s^k|$ and $l(\cdot, \cdot)$ are the number of instances in scenario $k$ and the cross-entropy loss, respectively.

Based on the above formulation, we can further define the multi-scenario multi-task information flow selection problem.
In MSMTR methods, to effectively capture the complex relationships between scenarios and tasks, the network structure is usually abstracted into four types of information units: the scenario-shared network $f_s^{sh}(\cdot)$, scenario-specific network $f_s^{k}(\cdot)$, task-shared network $f_t^{sh}(\cdot)$, and task-specific network $f_t^{m}(\cdot)$.
Thus, modeling the complete relationship information between the scenario and the task for each instance $\mathbf{x}_i^k$ can be formulated as:
\begin{equation}\label{eq:2}
    \mathcal{Q}_i^{{k},m} = f_t^{sh}\Big(f_s^{sh}(e_i^k) \oplus f_s^{k}(e_i^k)\Big) \oplus f_t^{m}\Big(f_s^{sh}(e_i^k) \oplus f_s^{k}(e_i^k)\Big), 
\end{equation}
where $e_i^k$ is the feature embedding of $\mathbf{x}_i^k$, $\oplus$ denotes the information fusion operation, and $\mathcal{Q}_i^{{k},m}$ is the complete relationship information integrating scenario and task signals.
However, as shown in Fig.~\ref{fig:intro}, interactions among the four information units generate four distinct relationship information flows between scenarios and tasks: scenario \& task-shared information $h_i^{sh,sh}$, scenario-shared \& task-specific information $h_i^{sh,m}$, scenario-specific \& task-shared information $h_i^{k,sh}$, and scenario \& task-specific information $h_i^{k,m}$.
These four relationship information flows can be viewed as the decomposition of the complex relationship information, each capturing a local dependency, and their fusion is expressed as:
\begin{equation}\label{eq:3}
    \mathcal{Q}_i^{{k},m} = h_i^{sh,sh} \oplus h_i^{sh,m} \oplus h_i^{k,sh} \oplus h_i^{k,m},
\end{equation}

Previous works generally focus on modeling the complete relationship information, as in Eq.~\eqref{eq:2}, but do not distinguish the varying importance of the four types of relationship information and simply integrate them all as in Eq.~\eqref{eq:3}.
Intuitively, different local relationships contribute differently to model performance.
For example, when modeling a sub-task within the main scenario, $h_i^{k,m}$ may be more critical, while other information may even act as noise and interfere with model learning. 
Hence, the multi-scenario multi-task information flow selection problem can be defined as learning a mask operation $\mathbf{G}_i$ to select unhelpful local information from the four flow types and prune it:
\begin{equation}\label{eq:4}
     \hat{\mathcal{I}}_i^{k,m} = \mathbf{G}_i \odot \mathcal{I}_i^{k,m}, \; \mathcal{I}_i^{k, m} = [h_i^{sh,sh}, h_i^{sh,m}, h_i^{k,sh}, h_i^{k,m}],
\end{equation}
where $\mathcal{I}_i^{k, m}$ is the information flow set and $\mathbf{G}_i \in \{0,1\}^{4}$, $\odot$ denotes element-wise multiplication, and $\hat{\mathcal{I}}_i^{k,m}$ is the useless information flow set obtained after the mask operation.
Next, we can prune the information in $\hat{\mathcal{I}}_i^{k,m}$ and feed the refined representation into the output layer to get the prediction results, which is expressed as:
\begin{equation}\label{eq:5}
   y_i^{k,m} = f^{o}\Big(\delta(\mathcal{Q}_i^{{k},m}, \hat{\mathcal{I}}_i^{k,m})\Big),
\end{equation}
where $\delta(\cdot, \cdot)$ and $f^{o}(\cdot)$ are the pruning function and output layer network, respectively.
 Finally, with all these formulations, we can formulate our problem as follows,
\begin{equation}\label{eq:6}
    \min_{\Theta,\{\mathbf{G}_i\}} \mathcal{L}(\mathcal{D}).
\end{equation}
where $\Theta$ denotes the network parameters.

\section{The Proposed Framework}\label{sec:method}
\begin{figure*}[htbp]
    \centering
    \includegraphics[width=1.\textwidth]{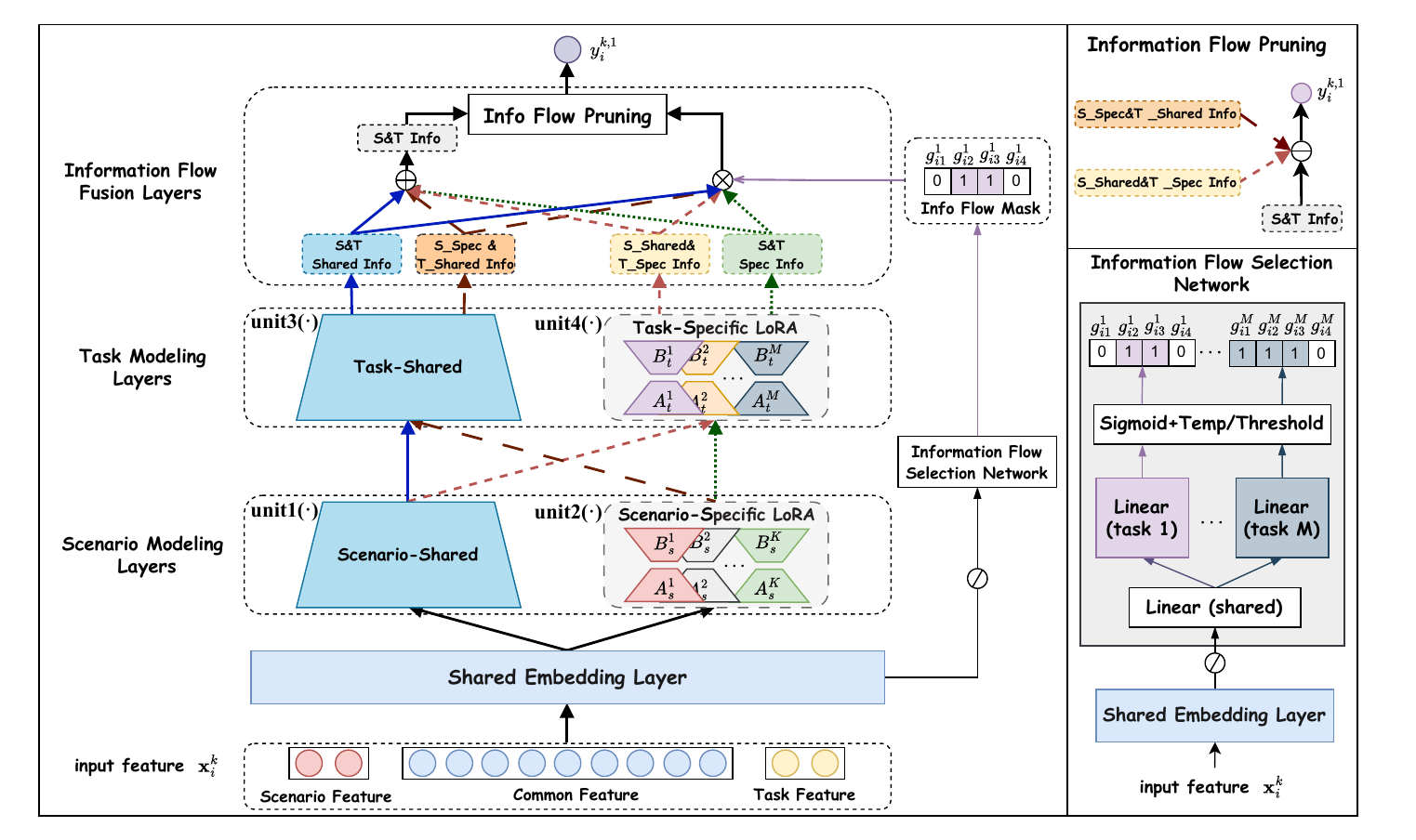}
    \Description[<Figure 2. Fully described in the text.>]{<A full description of Figure 2 can be found in Section 4.1.>}
    \caption{The architecture of the Automated Information Flow Selection (AutoIFS) framework, with Task 1 used as a specific task example, while the model actually outputs predictions for multiple tasks in parallel.}
    \label{fig:framework}
    \vspace{-10pt}
\end{figure*}
In this section, we first illustrate the overall framework of AutoIFS. Then, we detail each part of AutoIFS, including the learning and fusion of the four types of relationship information and the information flow selection. Finally, we introduce how to effectively optimize AutoIFS end-to-end based on MSMTR performance.
\subsection{Framework Overview}
The overall framework of AutoIFS is shown in Fig.~\ref{fig:framework}, which follows the multi-scenario multi-task modeling paradigm in Fig.~\ref{fig:intro}.
Our AutoIFS mainly consists of a hierarchical multi-scenario multi-task modeling network and an information flow selection network.
First, AutoIFS extracts and learns four types of relationship information between scenarios and tasks through four information units and introduces LoRA to simplify the fusion of relationship information.
Then, the information flow selection network dynamically learns masks for relationship information of different tasks, tailored to the instance input of a specific scenario.
Finally, through the information flow mask and information flow pruning, invalid relationship information is filtered out, and highlights the role of key relationship information, thereby improving model performance.
Notably, Fig.~\ref{fig:framework} uses task $1$ as an example to illustrate a specific task, while the model actually outputs predictions for multiple tasks in parallel.

\subsection{Information Flow Learning and Fusion}
According to our previous definition, we can directly use MoE or MLP networks as information units~\cite{hinet} and follow Eq.~\eqref{eq:2} to learn and fuse the four types of relationship information to model the complex dependencies between scenarios and tasks.
However, this is not an ideal solution.
Complex architectures complicate information fusion, reduce interpretability, and require substantial parameter growth as scenarios and tasks increase, raising training costs and making optimization more difficult.
Thus, we introduce low-rank adaptation (LoRA) \cite{LoRA, li2025toplora, li2025bora} to efficiently decouple the four information units without significantly increasing parameters, enabling more flexible and natural information fusion.
At the same time, inspired by previous studies~\cite{mlora, multilora, mola}, we can capture shared information through the full-rank parameter matrix $\mathbf{W}$ and use the low-rank network $\mathbf{BA}$ to learn specific information.
Hence, based on LoRA, we can realize four decoupled information units: $f_s^{sh}(\cdot)$, $f_s^{k}(\cdot)$, $f_t^{sh}(\cdot)$, and $f_t^{m}(\cdot)$, which are formulated as:
\begin{equation}\label{eq:9}
    f_s^{sh}(\mathbf{x}) = \mathbf{W}_s^{sh}\mathbf{x} + \mathbf{b}_s^{sh}, \; f_s^{k}(\mathbf{x}) = \mathbf{B}_s^{k}\mathbf{A}_s^{k}\mathbf{x} + \mathbf{b}_s^{k},
\end{equation}
\begin{equation}\label{eq:10}
    f_t^{sh}(\mathbf{x}) = \mathbf{W}_t^{sh}\mathbf{x} + \mathbf{b}_t^{sh}, \; f_t^{m}(\mathbf{x}) = \mathbf{B}_t^{m}\mathbf{A}_t^{m}\mathbf{x} + \mathbf{b}_t^{m}.
\end{equation}
where, $\mathbf{W}_s^{sh}$ and $\mathbf{W}_t^{sh}$ are the learnable parameter matrices for the scenario-shared and task-shared networks, $\mathbf{B}_s^{k}\mathbf{A}_s^{k}$ and $\mathbf{B}_t^{m}\mathbf{A}_t^{m}$ are the learnable low-rank matrices for the scenario-specific and task-specific networks, and $\mathbf{b}_s^{sh}$, $\mathbf{b}_s^{k}$, $\mathbf{b}_t^{sh}$, and $\mathbf{b}_t^{m}$ are the corresponding bias vectors.
Note that $\mathbf{A}_s^{k}$ and $\mathbf{A}_t^{m}$ are initialized with a Gaussian distribution and $\mathbf{B}_s^{k}$ and $\mathbf{B}_t^{m}$ are initialized with zero.
By combining Eq.~\eqref{eq:9} and Eq.~\eqref{eq:10}, we can make the learning and fusion of the four types of relationship information more flexible and straightforward, such as using addition, which is also one of the default settings of LoRA.
Thus, we reformulate Eq.~\eqref{eq:2} as:
\begin{equation}\label{eq:11}
    h_i^{sh, sh} = f_t^{sh}\Big(f_s^{sh}(e_i^k)\Big), \;  h_i^{sh,m} = f_t^{m}\Big(f_s^{sh}(e_i^k)\Big), 
\end{equation}
\begin{equation}\label{eq:12}
   h_i^{k, sh} = f_t^{sh}\Big(f_s^{k}(e_i^k)\Big), \; h_i^{k,m} = f_t^{m}\Big(f_s^{k}(e_i^k)\Big), 
\end{equation}
\begin{equation}\label{eq:13}
    \mathcal{Q}_i^{{k},m} = h_i^{sh,sh} + h_i^{sh,m} + h_i^{k,sh} + h_i^{k,m},
\end{equation}
Obviously, Eq.~\eqref{eq:11} and Eq.~\eqref{eq:12} independently learn four relationship information through the combination of information units and are completely decoupled from each other, which provides greater convenience for the subsequent Eq.~\eqref{eq:13} to use "+" fusion.
Notably, only when LoRA is used to implement information units and "+" fusion information is adopted, Eq.~\eqref{eq:11}, Eq.~\eqref{eq:12}, and Eq.~\eqref{eq:13} form the decomposed form of Eq.~\eqref{eq:2}.
Finally, since AutoIFS uses low-rank matrices to learn scenario-specific and task-specific information, this method effectively controls the increase in parameters, ensuring a lightweight model architecture.

\subsection{Task-Aware Information Flow Selection}
To represent discrete features and extract feature information, the model transforms each instance $\mathbf{x}{i}^k$ into a low-dimensional dense vector using a shared embedding layer:
\begin{equation}\label{eq:14}
    \mathbf{e}_{i}^{k}=\mathbf{E} \times\mathbf{x}_{i}^k,
\end{equation}
where $\mathbf{E}\in\Theta$ is the embedding table that is shared across scenarios.
Next, the feature embeddings of each instance are fed into the selection network, which aims to identify relationship information within $\mathcal{I}_i^{k, m}$ that negatively affects the MSMTR model’s ability to learn key preferences.
In other words, its function is to generate a task-wise probability vector for each instance and determine, via discretization, whether the corresponding  relationship information should be "removed" or "retained."
Although a single feed-forward network can serve as a selection module, prior work shows that such simple designs fail to capture differences across complex scenarios~\cite{ssim,multiemb}.
Considering that instances contain scenario attributes and the model must remain lightweight, we propose a task-aware information flow selection network with a hyper-network style structure to better highlight scenario-task relationships and make more reasonable selections of relationship information within tasks.

Specifically, we first perform shared feature extraction on $\mathbf{e}_{i}^{k}$ using a shared feed-forward neural network (FNN), where the $l$-th multi-layer perceptron (MLP) layer is defined as:

\begin{equation}\label{eq:15}
    \mathbf{h}^{l}_{i}=\sigma\left({W}^{l} \mathbf{h}^{l-1}_{i}+{b}^{l}\right), \quad l \in[1, L],
\end{equation}
where ${W}^{l}$ and ${b}^{l}$ represent the learnable weight matrix and bias vector, respectively, $\sigma(\cdot)$ indicates the $ReLU$ activation function, and $L$ is the number of MLP layers.
In the first MLP layer, $\mathbf{h}^{0}_{i}=\oslash(\mathbf{e}^{k}_{i})$, where $\oslash(\cdot )$ denotes truncated gradient backpropagation.
To fully capture task characteristics, we introduce task-specific feed-forward components inspired by hyper-networks. Each instance can generate the corresponding final output weight for each task through these components, and the relationship information weight vector is computed as:
\begin{equation}\label{eq:16}
    w_{i}^{m}=\mathbf{W}^{m} \mathbf{h}^{L}_{i}+\mathbf{b}^{m},
\end{equation}
where $w_{i}^{m}$ is the weight vector of relationship information in $i$-th instance. In addition, $\mathbf{W}^{m}$ and $\mathbf{b}^{m}$ denote the learnable task-specific weight matrix and bias vector of the $m$-th task.

Then, we need to normalize and approximately discretize $w_{i}^{m}$ to efficiently optimize the mask set $\{\mathbf{G}_i\}$ with instance-level granularity to obtain the continual gate vector for relationship information. 
Specifically, the continual gate $\mathbf{g}_{i}^{m}$ can be defined as,
\begin{equation}\label{eq:17}
    \mathbf{g}_{i}^{m}=\phi \left(w_{i}^{m} \times \tau\right), \quad \tau=\gamma^{p / P},
\end{equation}
where $\mathbf{g}_{i}^{m} \in \mathbf{G}_i$, $\phi(\cdot)$ is the sigmoid function, $p$ is the current training epoch, $P$ is the total training epoch and $\gamma$ is the final value of $\tau$ after training for $P$ epochs.
During training, by replacing the mask with continual gate vectors, we can optimize the parameters of the MSMTR model and the information flow selection network in a differentiable manner.
After training for $P$ epochs, the final gate vector $\mathbf{g}_{i}^{m}$ can be calculated using the unit-step function:
\begin{equation}
 \label{eq:18}
\mathbf{g}_{i}^{m}= \left\{\begin{array}{ll}
0, & w_{i}^{m} \leq 0, \\
1, & \text { otherwise. }
\end{array}\right.
\end{equation}

\subsection{Information Flow Pruning}
After obtaining $\mathbf{g}_{i}^{m}$ through Eq.~\eqref{eq:17} or Eq.~\eqref{eq:18}, we use Eq.~\eqref{eq:4} to determine the relationship information to be removed and then apply Eq.~\eqref{eq:5} for information pruning and final prediction.
However, the specific form of the information pruning function $\delta(\cdot, \cdot)$ is a crucial issue. We found that a simple subtraction operation can achieve good results. 
Its mathematical expression is as follows:
\begin{equation}\label{eq:19}
\begin{split}
\delta(\mathcal{Q}_i^{{k},m}, \hat{\mathcal{I}}_i^{k,m}) = &
    \mathcal{Q}_i^{{k},m} - \mathbf{g}_{i1}^{m} \odot h_i^{sh,sh} - \mathbf{g}_{i2}^{m} \odot h_i^{sh,m}  \\
   & - \mathbf{g}_{i3}^{m} \odot h_i^{k,sh} 
    - \mathbf{g}_{i4}^{m} \odot h_i^{k,m},
\end{split}
\end{equation}
where $\odot$ denotes element-wise multiplication. And, $\delta(\cdot,\cdot)$ can also be implemented in other ways as needed. In our experiments, we observed that $f^{o}$ actually weakens scenario-task relational characteristics, so AutoIFS removes this additional output layer network.

\subsection{Optimization}
This subsection will present the optimization process of our method, which consists of two stages: training and reuse.
\subsubsection{Training.}
At this stage, we introduce the optimization constraints necessary to ensure that the MSMTR model effectively learns the desired information mask.
To promote sparsity in the information mask vector and prevent excessive information pruning, we incorporate $l_1$ regularization.
 \begin{equation}
 \label{eq:sp}
    \mathcal{L}^{sp} =   \sum_{i=1}^{n} \sum_{m=1}^{M} \left \|  \mathbf{g}_{i}^{m} \right \|_{1},
\end{equation}
where $\left \| \cdot \right \|_{1}$ denotes the $l_{1}$ norm. 
Previous research~\cite{optfs} employed the $l_1$ norm as an approximation of the $l_0$ norm.
Building upon this approach and integrating Eq.\eqref{eq:1} and Eq.\eqref{eq:sp}, the final training objective is defined as follows:
\begin{equation}
\label{eq:loss_search}
\min_{\{\mathbf{G}_i\}, \Theta} \mathcal{L}_{AutoIFS}=\mathcal{L} + \lambda \mathcal{L}^{sp},
\end{equation}
where $\lambda$ denotes the regularization coefficient for sparsity control.

\subsubsection{Reuse.}
Since all potential relational information is incorporated into the model at the beginning of the training to identify the optimal set of information flow masks ${\mathbf{G}_i}$, some irrelevant or redundant relationships may be included, which could impair the model’s performance.
Therefore, we can retrain the model using the optimal ${\mathbf{G}_i}$ obtained to maximize performance.
Specifically, leveraging the well-trained information flow selection network, we retrain the model parameters $\Theta$ for $P_c$ epochs, where $P_c$ is a predefined hyperparameter. 
The final model parameters $\Theta$ can be optimized by Eq.\eqref{eq:1} without additional regularization as follows:
\begin{equation}
\label{eq:loss_retrain}
\min_{\Theta} \mathcal{L}.
\end{equation}

\subsection{Complexity Analysis}
To demonstrate the efficiency and lightweight nature of AutoIFS, we provide a brief complexity analysis.
AutoIFS comprises two key components: LoRA modules, which model four types of relationship information flow, and an information flow selection network, which identifies the most relevant flows.  
First, in the backbone network, LoRA replaces the original MLP or MoE structures with low-rank adapters, reducing the number of parameters while maintaining modeling capacity. 
Second, the selection network is a compact MLP that determines the retention of each information flow based on input features, without involving any complex computation; thus, its additional computational overhead is negligible.
Overall, compared with the original MSMTR model, AutoIFS introduces only minimal extra parameters and FLOPs, while also benefiting from LoRA’s low-rank structure to reduce parameter size. 
As reported in Section 5.4, it achieves comparable training and inference costs while substantially reducing model size, striking a favorable balance between computational efficiency and predictive performance.
\section{Experiments}\label{sec:experiments}
In this section, we conduct a series of experiments to answer the following five critical questions.
Note that the source codes are available at \url{https://github.com/ChaohuaYang/AutoIFS}.
\begin{itemize}[leftmargin=*]
    \item RQ1: How does AutoIFS perform compared to the baselines?
    \item RQ2: What is the role of some key components in AutoIFS?
    \item RQ3: How efficient is AutoIFS compared to other baselines?
    \item RQ4: What are the effects of different settings of key parameters and the information flow mask on AutoIFS?
    \item RQ5: How does AutoIFS perform in real MSMTR scenarios?
\end{itemize}

\subsection{Experiment Setup}\label{sec:experiments:setup}
\subsubsection{Datasets.}
We evaluate the efficacy of AutoIFS on two public benchmark recommendation datasets: MovieLens-1M\footnote{https://grouplens.org/datasets/movielens/} and KuaiRand-Pure\footnote{https://kuairand.com/}. 
MovieLens-1M is a movie rating dataset provided by \textit{GroupLens Research}. It contains approximately 1 million ratings for 3,900 movies, along with seven user attributes and 2 item attributes. 
The dataset includes diverse information such as ratings, user details across different scenarios and tasks, and demographic data like gender, age, and occupation.
KuaiRand-Pure has collected recommendation logs from the video-sharing mobile application Kuaishou.
It contains more than two million interactions on 7,583 videos and includes 30 user features and 62 item features.

\subsubsection{Dataset Preprocessing}
\label{sec:preprocessing}
For MovieLens-1M, following the setting of previous work~\cite{m3oe}, we use the feature "age" to split the dataset into three scenarios and infer two tasks: "click" and "like".
Since the original dataset only provides ratings from 1 to 5, we set the rating thresholds to "4" and "5" respectively to obtain the labels of the two tasks: $rating \geq 4$ indicates that the user clicks, and $rating \geq 5$ indicates that the user likes.
Then, since there are fewer non-ID features in MovieLens-1M, we no longer filter low-frequency features and divide the training, validation, and test sets from all the raw data of each scenario, containing the ratios of 80\%, 10\%, and 10\%.
For KuaiRand-Pure, following the settings of previous work~\cite{m3oe, optembed}, we select "tab" representing interactions on different tabs to divide the dataset into 3 scenarios and address 2 tasks: "click" and "long-view".  
Then, we filter low-frequency features using a threshold of 2 and divide the training, validation, and test sets from all the raw data of each scenario,  containing the ratios of 80\%, 10\%, and 10\%.
The processed datasets' statistics are shown in Table~\ref{tab:datasets}.
\begin{table}
\centering
\caption{Dataset statistics.}
\vspace{-5pt}
\resizebox{0.48\textwidth}{!}{
\begin{tabular}{c|ccc|ccc}
\specialrule{0.1em}{1pt}{1pt}
        \multicolumn{1}{c|}{Dataset} & \multicolumn{3}{c|}{MovieLens-1M} & \multicolumn{3}{c}{KuaiRand-Pure} \\
    \cline{2-7}
            \multicolumn{1}{c|}{Scenario} & S1 & S2 & S3 & S1 & S2 & S3 \\
    \hline
        \#Users & 1,325 & 2,096 & 2,619 & 15,398 & 27,049 & 11,809 \\
        \#Items & 3,429 & 3,508 & 3,595 & 6,233 & 7,580 & 4,633  \\
        \#Instances & 210,747 & 395,556 & 393,906 & 178,087 & 2,236,414 & 93,165  \\
        Percentage & 21.07\% & 39.55\% & 39.38\% & 7.10\% & 89.18\% & 3.72\%  \\
\specialrule{0.1em}{1pt}{1pt}
\end{tabular}}
\label{tab:datasets}
\vspace{-10pt}
\end{table}

\subsubsection{Metrics.}
Following the setup of previous works~\cite{m3oe, MultiFS}, we use two key evaluation metrics widely used in deep recommender systems: area under the ROC curve (AUC) and cross-entropy (log loss). 
Notably, an improvement of more than 0.1\% in AUC is considered significant for the CTR prediction task~\cite{DeepFM}.

\begin{table*}[htbp]
\centering
\caption{Results on all datasets, where the best and second best results are marked in bold and underlined, respectively. Note that $^{*}$ indicates a significance level of $p\leq 0.05$ based on a two-sample t-test between our method and the best baseline.}
\vspace{-5pt}
\resizebox{1.0\textwidth}{!}{
\begin{tabular}{c|cccccc|cccccc||cccc}
\specialrule{0.1em}{1pt}{1pt}
    \multirow{3}{*}{Dataset}  & \multicolumn{12}{c||}{AUC for Each Scenario and Task} & \multicolumn{4}{c}{Overall Performance} \\
    \cline{2-17}
    & \multicolumn{6}{c|}{MovieLens-1M} & \multicolumn{6}{c||}{KuaiRand-Pure} & \multicolumn{2}{c}{MovieLens-1M} & \multicolumn{2}{|c}{KuaiRand-Pure} \\
    \cline{2-17}
    & S1, T1 & S1, T2 & S2, T1 & S2, T2 & S3, T1 & S3, T2 & S1, T1 & S1, T2 & S2, T1 & S2, T2 & S3, T1 & S3, T2 & \multicolumn{1}{c}{AUC$\uparrow$} & \multicolumn{1}{c}{Logloss$\downarrow$} & \multicolumn{1}{c}{AUC$\uparrow$} & \multicolumn{1}{c}{Logloss$\downarrow$} \\
    \hline
        DNN-T & .8134 & .8318 & .8256 & .8292 & .8103 & .8161 & \underline{.7055} & .7160 & .7843 & .7971 & .7205 & .7222 & .8211 & .4626 & .7409 & .4408 \\
        MMoE-T & .8120 & .8313 & .8232 & .8289 & .8086 & .8155 & .6944 & .7064 & .7853 & .7999 & .7227 & .7264 & .8199 & .6256 & .7392 & .4845 \\
        MLoRA-T & .8143 & .8325 & .8257 & .8312 & .8099 & .8153 & .6938 & .7112 & .7850 & .7989 & .7266 & .7288 & .8214 & .4665 & .7407 & .4368 \\
        \hline
        DNN-S & .8136 & .8324 & .8248 & .8301 & .8090  & .8158 & .6996 & \underline{.7193} & .7832 & .7978 & .7252 & .7295 & .8210 & .4651 & .7424 & .4343 \\
        MMoE-S & .8113 & .8301 & .8235 & .8286 & .8082 & .8163 & .6943 & .7049 & .7851 & .7998 & .7206 & .7243 & .8197 & .5565 & .7381 & .4876 \\
        MLoRA-S & .8151 & .8329 & .8254 & .8315 & .8100 & \underline{.8173} & .7034 & .7133  & .7861 & .8003 & .7241 & .7272 & .8220 & .4623 & .7424 & .4373 \\
        \hline
        DNN-ST & .8126 & .8308 & .8250  & .8278 & .8106 & .8146 & \textbf{.7101} & .7178 & .7796 & .7941 & .7270 & .7306 & .8202 & .4662 & .7432 & .4339 \\
        MMoE-ST & .8154 & \underline{.8337} & .8260  & \underline{.8318} & \underline{.8113} & .8169 & .7031 & .7144 & .7831 & .7984 & .7232 & .7258 & \underline{.8225} &  \underline{.4614} & .7413 & .4373 \\
        MLoRA-ST & .8140 & .8324 & .8261 & .8314 & .8097 & .8172 & .7004 & .7155 & .7863 & .8003 & .7289 &  \underline{.7333} & .8218 & .4624 & .7441 &  \underline{.4329} \\
        M3oE & .8142 & .8333 & .8253 & .8301 & .8092 & .8151 & .6975 & .7155 & .7847 & .7990  & .7272 & .7302 & .8212 & .4651 & .7423 & .4360 \\
        M2M & .8160  & .8329 & .8259 & .8304 & .8105 & .8167 & .6983 & .7084 & \underline{.7880}  & \underline{.8020}  & .7274 & .7305 & .8221 & .4675 & .7424 & .4354 \\
        HiNet & \underline{.8161} & .8330  & \underline{.8262} & .8314 & .8100   & .8160  & .7010  & .7174 & .7869 & .8013 & \underline{.7290}  & .7328 & .8221 & \underline{.4614} & \underline{.7447} & .4347 \\
        \hline
        AutoIFS & $\textbf{.8191}^{*}$ & $\textbf{.8361}^{*}$ & $\textbf{.8282}^{*}$ & $\textbf{.8351}^{*}$ & $\textbf{.8134}^{*}$ & $\textbf{.8196}^{*}$ & .7020  & $\textbf{.7218}^{*}$ & \textbf{.7885} & \textbf{.8023} & $\textbf{.7320}^{*}$  & $\textbf{.7357}^{*}$ & $\textbf{.8253}^{*}$ & $\textbf{.4575}^{*}$ & $\textbf{.7471}^{*}$ & \textbf{.4319} \\
    \specialrule{0.1em}{1pt}{1pt}
    
\end{tabular}}
\label{tab:main_result}
\vspace{-10pt}
\end{table*}
\subsubsection{Baseline Methods.}
To examine the performance of our approach, we selected representative methods from multi-task recommendation, multi-scenario recommendation, and multi-scenario multi-task recommendation for comparison, i.e., DNN, MMoE~\cite{mmoe}, MLoRA~\cite{mlora}, M3oE~\cite{m3oe}, M2M~\cite{m2m}, and HiNet~\cite{hinet}. 
Here, for a more comprehensive comparison, we implement the multi-task, multi-scenario, and multi-task multi-scenario versions of DNN, MMoE, and MLoRA, respectively. Specifically, we denote the suffix "-T" for the multi-task setting, the suffix "-S" for the multi-scenario setting, and the suffix "-ST" for the multi-scenario multi-task setting.

\subsubsection{Implementation Details.}
This subsection details the implementation of our AutoIFS and the baseline methods. 
For the general hyperparameters, the embedding dimension and batch size are 16 and 4096, respectively.
Following previous work~\cite{optfs}, we select the optimal learning rate from $\{$1e-3, 3e-4, 1e-4, 3e-5, 1e-5$\}$ and the $l_2$ regularization from $\{$1e-3, 3e-4, 1e-4, 3e-5, 1e-5, 3e-6, 1e-6$\}$, and we employ the Adam optimizer along with Xavier initialization for the experiments.
We implement the MLP layers in the baseline models as a three-layer fully connected network with dimensions $[1024, 512, 256]$.
For the hyperparameters of AutoIFS, we select the optimal rank $r$ and final value $\gamma$ from $\{$2, 4, 8, 16, 32, 64$\}$ and $\{$50, 100, 500, 1000, 5000, 10000$\}$, respectively.
And, we select the optimal sparse regularization penalty $\lambda$ from $\{$0, 1e-3, 5e-3, 1e-2, 5e-2, 1e-1, 5e-1$\}$, training epochs $P$ from $\{$5, 10, 15, 20$\}$. 
During the reuse stage, we reselect the optimal learning ratio and $l_2$ regularization and
choose the rewinding epoch $P_c$ from $\{$1, 2, $\cdots$, $P-1$$\}$.
For other baseline methods, we use the open source implementations for MMoE~\cite{mmoe}, M3oE~\cite{m3oe} and MLoRA~\cite{mlora}.
In addition, we re-implemented HiNet with reference to the open source repositories of HiNet~\cite{hinet}.
Since the implementations for M2M~\cite{m2m} are not publicly available, we have carefully re-implemented the method following the descriptions and details outlined in the original paper.
Note that all baseline models have been carefully tuned to achieve their optimal performance by exploring a consistent range of hyperparameter settings. 

\begin{table*}[htbp]
\centering
\caption{Ablation Analysis on our AutoIFS, where the best results are marked in bold.}
\vspace{-5pt}
\resizebox{1.0\textwidth}{!}{
\begin{tabular}{c|cccccc|cccccc}
\specialrule{0.1em}{1pt}{1pt}
    \multirow{3}{*}{Dataset}  & \multicolumn{12}{c}{AUC for Each Scenario and Task} \\
    \cline{2-13}
    & \multicolumn{6}{c|}{MovieLens-1M} & \multicolumn{6}{c}{KuaiRand-Pure} \\
    \cline{2-13}
    & S1, T1 & S1, T2 & S2, T1 & S2, T2 & S3, T1 & S3, T2 & S1, T1 & S1, T2 & S2, T1 & S2, T2 & S3, T1 & S3, T2  \\
    \hline
        n.re. & 0.8171 & 0.8345 & 0.8269 & 0.8332 & 0.8117 & 0.818  & 0.7000  & 0.7206 & 0.7876 & 0.8011 & 0.7275 & 0.7309 \\
        n.sn. & 0.8167 & 0.8349 & 0.8268 & 0.8329 & 0.8115 & 0.8179 & \textbf{0.7035} & 0.7163 & 0.7863 & 0.8011 & 0.7295 & 0.7312 \\
        n.di. & 0.8168 & 0.8350  & 0.8272 & 0.8336 & 0.8125 & 0.8187 & 0.7013 & 0.7200   & 0.7871 & 0.8019 & 0.7288 & 0.7328 \\
        n.rp. & 0.8191 & 0.8361 & 0.8282 & 0.8351 & 0.8134 & 0.8196 & 0.7016 & 0.7196 & 0.7880  & 0.8021 & 0.7297 & 0.7346 \\
        w.rs. & 0.8177 & 0.8349 & 0.8264 & 0.8326 & 0.8113 & 0.8173 & 0.6963 & 0.7170  & 0.7867 & 0.8015 & 0.7269 & 0.7301 \\
        w.fn. & 0.8167 & 0.8344 & 0.8269 & 0.8325 & 0.8121 & 0.8180  & 0.6976 & 0.7163 & 0.7870  & 0.8020  & 0.7273 & 0.7306 \\
        w.me. & 0.8169 & 0.8356 & 0.8245 & 0.8245 & 0.8097 & 0.8177 & 0.6929 & 0.7143 & 0.7856  & 0.8012 & 0.7214 & 0.7332 \\
        w.pt. & 0.8169 & 0.8345 & 0.8271 & 0.8335 & 0.8124 & 0.8191 & 0.7007 & 0.7141 & 0.7880  & 0.8022 & 0.7315 & 0.7345 \\
        AutoIFS     & \textbf{0.8191} & \textbf{0.8361} & \textbf{0.8282} & \textbf{0.8351} & \textbf{0.8134} & \textbf{0.8196} & 0.7020 & \textbf{0.7218} & \textbf{0.7885} & \textbf{0.8023} & \textbf{0.7320} & \textbf{0.7357} \\
    \specialrule{0.1em}{1pt}{1pt}
\end{tabular}}
\label{tab:ablation_result}
\vspace{-10pt}
\end{table*}
\subsection{RQ1: Overall Performance}\label{sec:experiments:rq1}

In this subsection, we compare AutoIFS with three types of baselines: multi-task, multi-scenario, and multi-scenario multi-task recommendations.
The overall performance of AutoIFS and other baseline methods is presented in Table~\ref{tab:main_result}. 

From Table~\ref{tab:main_result}, the following observations can be made:
1) The performance of multi-task methods and multi-scenario methods on both datasets is generally inferior to that of multi-scenario multi-task methods. 
This indicates that in complex applications with multiple scenarios and tasks, relying solely on multi-task or multi-scenario learning cannot capture the complex interactions between scenarios and tasks, resulting in mediocre model performance.
2) Multi-scenario multi-task methods redesigned based on existing models, i.e., DNN-ST, MMoE-ST, MLoRA-ST, slightly outperform multi-task or multi-scenario methods at first glance, but show unstable performance across two datasets.
For instance, from the perspective of average AUC, DNN-ST performs relatively poorly on MovieLens-1M but performs well on Kuairand-Pure.
This means that simply modified multi-scenario multi-task modeling fails to fully capture the relationships between scenarios and tasks.
3) Multi-scenario multi-task methods that focus on architectural design, i.e., M3oE, M2M, and HiNet, have relatively stable overall performance, especially HiNet, which has an average performance that is generally better than other baselines. 
This proves that customized complex architecture design can indeed better model the complex relationships between scenarios and tasks.
4) Low-rank adaptation methods like MLoRA-T, MLoRA-S, and MLoRA-ST perform on par with similar methods while reducing model parameters, demonstrating that low-rank modules can further unlock the model's potential in complex environments.
5) Our AutoIFS performs significantly better than all baseline methods on both datasets, which proves that AutoIFS can effectively model complex relationships between scenarios and tasks. 
Furthermore, this validates that AutoIFS can alleviate negative information transfer between scenarios and tasks by introducing the information flow selection network to prune useless relationship information.

\subsection{RQ2: Ablation Study}\label{sec:experiments:rq2}

In this subsection, we perform an ablation study to evaluate the impact of key steps and components in AutoIFS. 
Specifically, we first sequentially remove the reuse step (denoted as `n.re.'), the information flow selection network (denoted as `n.sn.'), the discretization operation (denoted as `n.di.'), and the regularization penalty (denoted as `n.rp.') from AutoIFS.
Then, we consider modifying the information flow selection network structure in AutoIFS to random selection (denoted as `w.rs.') and a single feedforward neural network (denoted as `w.fn.').
Additionally, we modified the backbone network in AutoIFS by replacing the two LoRA networks with MoE networks (denoted as `w.me.') and adding prediction towers for the two tasks (denoted as `w.pt.').
We report the corresponding results in Table~\ref{tab:ablation_result}.
We can observe that removing the reuse step usually leads to performance degradation, indicating that it is essential to mitigate the impact of jointly optimizing the model network and the information flow selection network.
Removing the information flow selection network or the discretization operation causes performance loss, highlighting their benefits in AutoIFS.
Removing the regularization penalty term, that is, setting $\lambda$ to 0, keeps performance on MovieLens-1M unchanged but degrades AutoIFS performance on Kuairand-Pure, which shows that the regularization penalty term is also necessary.
Replacing the information flow selection network with a random selection mechanism or a feedforward neural network leads to performance degradation, emphasizing the importance of introducing a learnable information flow selection network.
Replacing the LoRA network with an MoE network also leads to degraded performance, confirming our earlier claim that LoRA is more effective for modeling relationship information.
Adding prediction towers for two tasks in AutoIFS decreases performance, possibly because they focus on modeling homogeneous information, which weakens the ability to capture specific scenario-task relationships.
This is also the reason why we do not introduce additional prediction towers in AutoIFS, and this design can further reduce model parameters.

\subsection{RQ3: Efficiency Analysis}\label{sec:experiments:rq3}
This subsection aims to validate the efficiency advantage of AutoIFS by comparing its model size and training/inference time against other baseline methods.
\begin{figure}[htbp]
\centering  \includegraphics[width=1.0\linewidth]{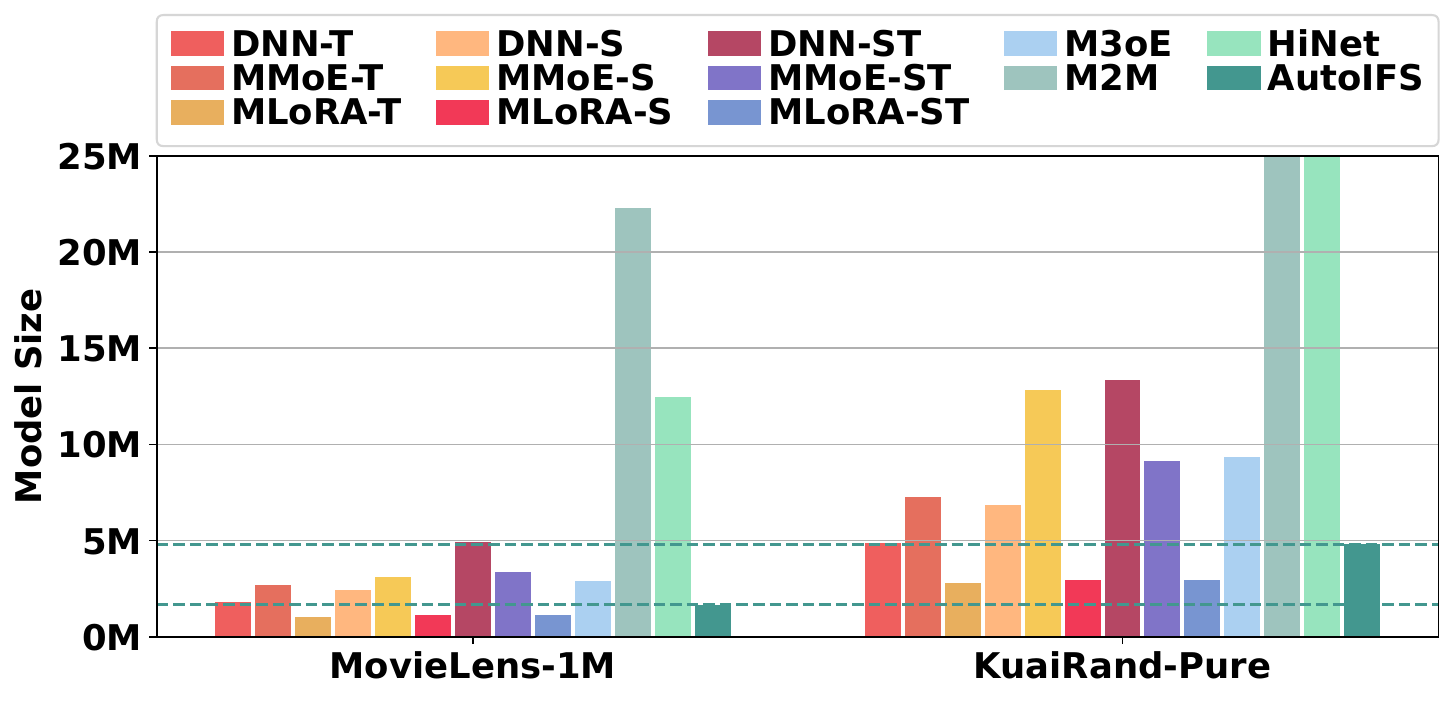}
\Description[<Figure 3. Fully described in the text.>]{<A full description of Figure 3 can be found in the first paragraph of Section 5.4.>}
\caption{Model size of our AutoIFS and baseline models on two datasets.}
\label{fig:size}
\vspace{-10pt}
\end{figure}
On the one hand, in actual industrial platforms, resources are usually limited, so storage-friendly recommendation models are easier to train and deploy. We show the model size of our AutoIFS and other baseline methods on two datasets in Fig.~\ref{fig:size}.
We can observe that the number of parameters AutoIFS has is only slightly higher than those of low-rank models, i.e., MLoRA-T, MLoRA-S, and MLoRA-ST. It is comparable to DNN-T but significantly lower than that of other methods.
However, according to the results in Table~\ref{tab:main_result}, we find that although MLoRA-T, MLoRA-S, MLoRA-ST, and DNN-T have fewer parameters, they fail to model the complex relationship between scenarios and tasks effectively.
In addition, MSMTR models like M3oE, M2M, and HiNet adopt complex architectures, which introduce substantial parameter overhead and may lead to suboptimal performance due to training difficulties.
In contrast, our AutoIFS introduces low-rank modules to flexibly model the relationships between scenarios and tasks while maintaining a compact model size.

On the other hand, training and inference time also affect the practical applicability of the model.
We report the per-epoch training time and test-set inference time for AutoIFS and other MSMTR baselines on the MovieLens-1M dataset in Table~\ref{tab:times}.
The results show that AutoIFS has training and inference times comparable to DNN-ST and MLoRA-ST, and significantly lower than M3oE, M2M, and HiNet.
This demonstrates that AutoIFS is an efficient MSMTR method and also validates our earlier claim that the additional computational overhead introduced by the information flow selection network is negligible.

\begin{table}[htbp]
\centering
\caption{Training and inference time statistics on the MovieLens-1M dataset.}
\vspace{-5pt}
\resizebox{1.0\linewidth}{!}{
\begin{tabular}{c|ccccccc}
\specialrule{0.1em}{1pt}{1pt}
        Times (s) & DNN-ST & MMoE-ST & MLoRA-ST & M3oE & M2M & HiNet & AutoIFS  \\
    \hline
        Train (Epoch) & 10.55 & 11.41 & 10.23 & 12.12 & 30.30 & 18.47 & 10.66  \\
        Inference & 0.0038 & 0.0288 & 0.0033 & 0.0605 & 0.1978 & 0.0774 & 0.0039 \\
\specialrule{0.1em}{1pt}{1pt}
\end{tabular}}
\label{tab:times}
\vspace{-10pt}
\end{table}

\subsection{RQ4: In-depth Analysis of AutoIFS}\label{sec:experiments:rq4}
Next, we conduct an analysis of the sensitivity of key parameters in AutoIFS and the interpretability of its information flow mask.

\begin{figure}[htbp] 
\centering
\includegraphics[width=1.0\linewidth]{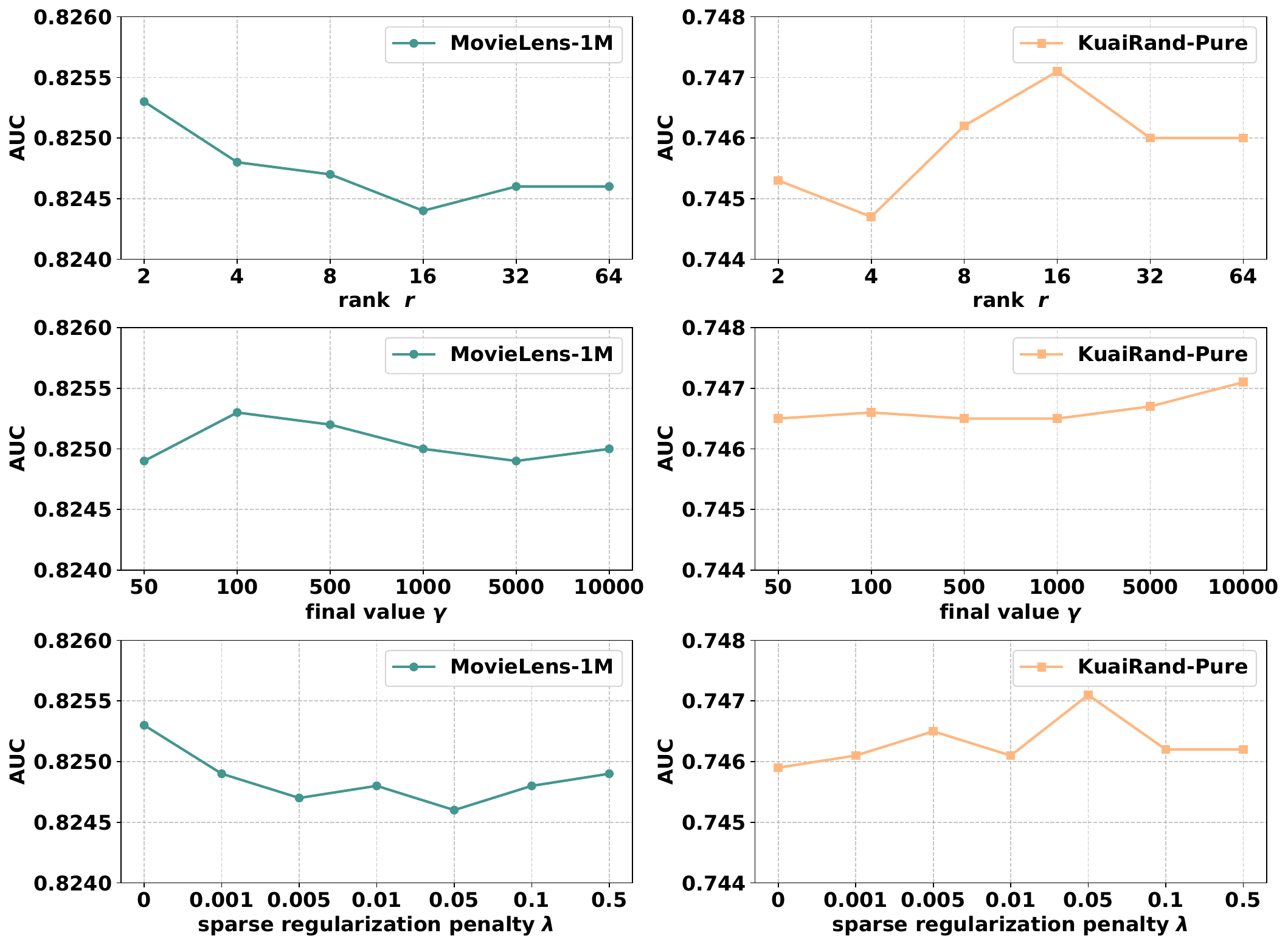}
\Description[<Figure 4. Fully described in the text.>]{<A full description of Figure 4 can be found in the second paragraph of Section 5.5.>}    
\caption{Sensitivity analysis of the rank $r$, the final value $\gamma$ and the regularization penalty $\lambda$ on two datasets.}
\label{fig:hyper}
\vspace{-5pt}
\end{figure}

\noindent\textbf{Parameter Sensitivity.}
Our AutoIFS introduces the low-rank module, in which the rank size $r$ is a crucial hyperparameter. The rank size $r$ directly affects the model's parameter scale and significantly impacts its performance. 
Moreover, the final value $\gamma$ and the sparsity regularization penalty $\lambda$ are two key parameters for mask learning of the information flow selection network. 
To further study the impact of these three key parameters on the performance of AutoIFS under different settings, we conducted experiments on two datasets, and the results are shown in Fig.~\ref{fig:hyper}.
As shown in the top two sub-figures in Fig.~\ref{fig:hyper}, on the MovieLens-1M, AutoIFS achieves the best performance when $r=2$. However, as $r$ increases, the model's performance declines significantly. 
In contrast, on the Kuairand-Pure, the model performs poorly when $r$ is small but achieves optimal performance when $r=16$. 
This difference may be due to the smaller data size of the MovieLens-1M, where a lower rank helps prevent overfitting. Unlike the Kuairand-Pure, which requires a higher rank to capture richer information, thereby improving model performance.
Thus, we set $r$ of AutoIFS to 2 and 16 in MovieLens-1M and Kuairand-Pure, respectively.
In addition, we also observe that $\gamma$ and $\lambda$ lead to only slight variations in overall performance under different settings, and all configurations significantly outperform the baseline methods.
This shows that the mask operation we designed maintains good stability and practicality.

\begin{figure}[htbp] 
\centering
\includegraphics[width=1.0\linewidth]{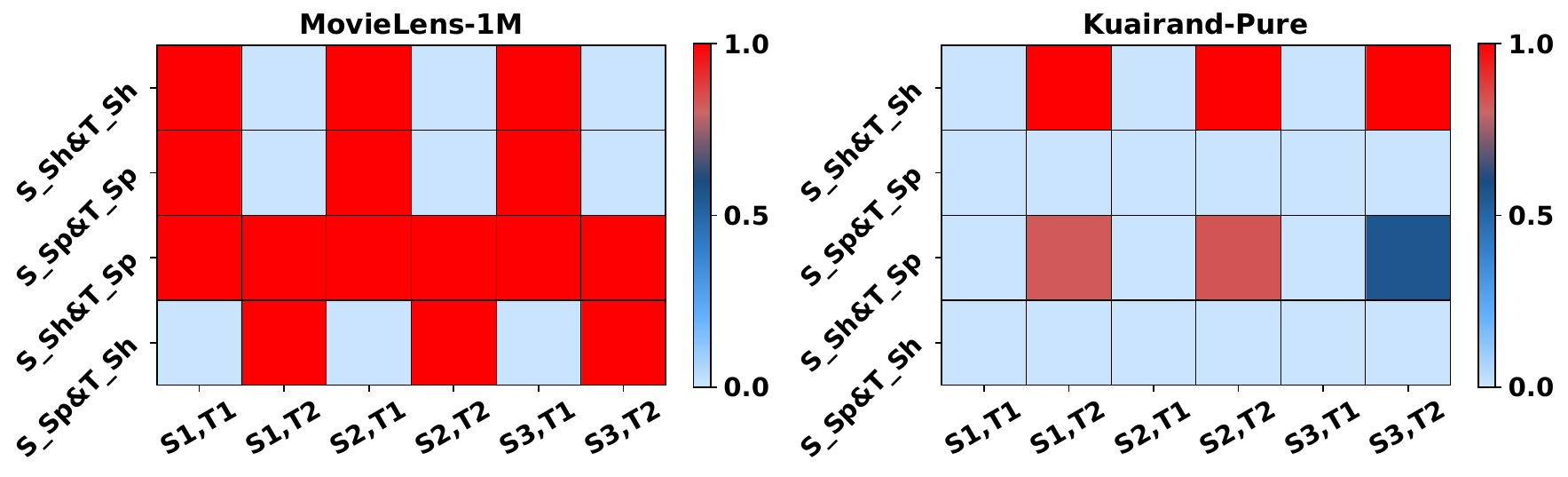}
\Description[<Figure 5. Fully described in the text.>]{<A full description of Figure 5 can be found in the third paragraph of Section 5.5.>} 
\caption{Mask Visualization of our AutoIFS on two datasets.}
\label{fig:gate}
\vspace{-10pt}
\end{figure}

\noindent\textbf{Interpretability of the Information Flow Mask.}
Intuitively, different information interactions between scenarios and tasks contribute differently to the model. 
This insight is at the core of our design of the information flow selection network, which is used to dynamically regulate the impact of different relationship information between scenarios and tasks so as to more accurately capture their key role and to improve interpretability by highlighting which relationships are most influential in different contexts.
After the information flow selection network is trained, we analyze the pruning ratio of the four relationship information on each scenario and task in the reusing stage, and the results are shown in Fig.~\ref{fig:gate}.
We can see that on MovieLens-1M, for instance, task 1 tends to prune scenario-shared \& task-shared relationship information and scenario-specific \& task-specific relationship information. In contrast, task 2 primarily prunes scenario-shared \& task-specific relationship information and scenario-specific \& task-shared relationship information.
However, on Kuairand-pure, all relationship information must be retained on task 1, while scenario-shared \& task-shared relationship information and scenario-shared \& task-specific relationship information are not favored on task 2.
This illustrates that the information flow mask can provide interpretability by explicitly revealing the varying importance of different relationship types for each task and scenario.

\subsection{RQ5: Online Experiments}\label{subsec:online}
\begin{figure}[htbp]
\includegraphics[width=1.\linewidth]{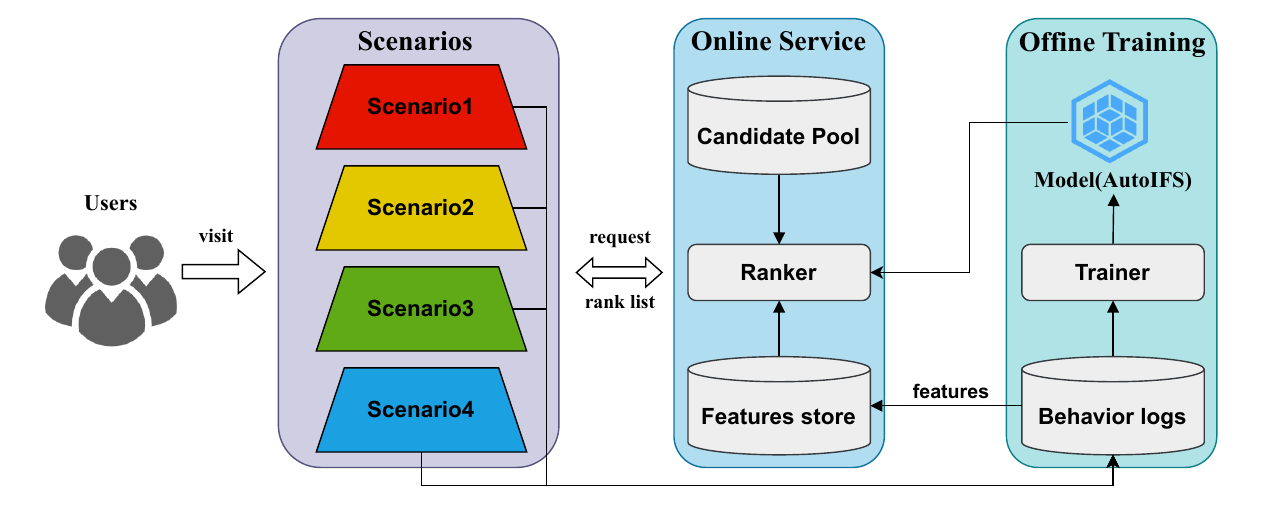}
\Description[<Figure 6. Fully described in the text.>]{<A full description of Figure 6 can be found in Section 5.6.>}
\caption{Overview of financial MSR scenarios.}
\label{fig:online}
\vspace{-15pt}
\end{figure}

Finally, to further evaluate the performance of our AutoIFS, we deploy it in a real recommendation business scenario of Tencent FiT, one of China's large online financial platforms.
The specific process of model deployment in the recommendation system is illustrated in Fig.~\ref{fig:online}.
Specifically, we randomly selected 10\% of users to use AutoIFS as the experimental group, and another 10\% of users to use the baseline model as the control group, and conducted an online A/B test for two consecutive weeks.
We used three key indicators commonly used in the platform (i.e., click-through rate (CTR), conversion rate (CVR), and subscription amount (SA)) as evaluation indicators.
As shown in Table~\ref{tab:online_result}, our AutoIFS achieved significant improvements in all indicators in all scenarios compared with the deployed baseline model, which also proves that AutoIFS effectively models more critical relationships between scenarios and tasks by pruning useless relationship information.
In addition, the average online inference latency is 2.74 ms for the baseline model and 2.98 ms for AutoIFS, which also shows that the information flow selection network introduced by our method has little impact on efficiency.

\begin{table}[htbp]
\centering
\caption{Gains obtained by AutoIFS in online deployments.}
\resizebox{1.0\linewidth}{!}{
\begin{tabular}{c|cccc|c}
\specialrule{0.1em}{1pt}{1pt}
        Metrics & Scenario1 & Scenario2 & Scenario3 & Scenario4 & Average   \\
    \hline
        CTR & 0.48\% & 1.02\% & 1.55\% & 0.43\% & 0.87\%   \\
        CVR & 0.94\% & 1.54\% & 4.56\% & 2.79\% & 2.46\%  \\
        SA & 11.77\% & 7.14\% & 5.54\% & 8.81\% & 8.32\%  \\
\specialrule{0.1em}{1pt}{1pt}
\end{tabular}}
\label{tab:online_result}
\vspace{-10pt}
\end{table}

\section{Conclusions}\label{sec:conclusions}
In this paper, we propose an Automated Information Flow Selection (AutoIFS) framework for multi-scenario multi-task recommendation. This framework enhances MSMTR performance by adaptively pruning irrelevant relationship information from the complex interactions between scenarios and tasks.
Specifically, our AutoIFS introduces low-rank adaptation technology to decouple and fuse relationship information more flexibly while minimizing increases in parameters and complexity.
Furthermore, we developed an information flow selection network to filter out irrelevant scenario-task relationship information, allowing key relationship information to play a more prominent role and thereby significantly enhancing model performance.
Finally, we conduct experiments on two public benchmark datasets to verify AutoIFS's effectiveness and analyze its unique properties. We also further demonstrate its gain in online real-world MSMTR scenarios.

\begin{acks}
We thank the support of the National Natural Science Foundation of China (Nos. 62302310, 62272315).
\end{acks}

\section*{Ethical Considerations}
Similar to existing recommendation methods, the proposed AutoIFS may reinforce users’ historical preferences, potentially affecting content diversity and information exposure. In practical deployment, such effects can be mitigated through system-level strategies, including debiasing and diversity-aware ranking mechanisms. Moreover, AutoIFS is developed and evaluated using publicly available anonymized datasets, thereby complying with established privacy and data protection standards.

\bibliographystyle{ACM-Reference-Format}
\bibliography{sample-base}

\end{document}